\documentclass[journal]{IEEEtran}

\usepackage{subfigure}
\usepackage{amsmath}
\usepackage{amssymb}
\usepackage{amsfonts}
\usepackage{latexsym}
\usepackage{xspace}
\usepackage{array}
\usepackage{epsfig}
\usepackage{times}
\usepackage{bm}
\usepackage{colortbl}
\usepackage{dsfont}
\usepackage{graphics}
\usepackage{graphicx}
\usepackage{multicol}
\usepackage{color}
\usepackage{epsfig}

\newcommand{\Hz}{\ensuremath\text{Hz}}

\ifCLASSINFOpdf

\else

\fi

\begin{document}

\title{Human brain distinctiveness based on EEG spectral coherence connectivity}

\author{Daria~La Rocca, Patrizio~Campisi, Balazs~Vegso, Peter~Cserti, Gyorgy~Kozmann, Fabio~Babiloni, \\ Fabrizio~De Vico Fallani.

\thanks{This work was in part supported by a grant
from the Italian Minister of Foreign Affairs "Direzione Generale per la
Promozione del sistema Paese" between Italy and Hungary. F. De Vico Fallani acknowledges the program ``Investissements d'avenir'' ANR-10-IAIHU-06.}%
\thanks{D. La Rocca and P. Campisi are with Section of Applied Electronics, Dept. Engineering,
Universit\`{a} degli Studi ``Roma Tre'', Roma, Italy.}
\thanks{B. Vegso, P. Cserti and G. Kozmann are with Dept. Electrical Engineering and Information Systems, Faculty of Information Technology,
University of Pannonia, Veszprem, Hungary.}%
\thanks{F. Babiloni is with IRCCS Fondazione Santa Lucia; Dept. Physiology and Pharmacology, University of Rome ``Sapienza''; BrainSigns srl; Rome, Italy.}
\thanks{F. De Vico Fallani (corresponding author) is with UPMC Univ Paris 06, UM75; Inserm, U1127; CNRS, UMR7225; ICM; Inria, Aramis project-team; Paris, France. Email fabrizio.devicofallani@gmail.com.}}%

\maketitle

\begin{abstract}
The use of EEG biometrics, for the purpose of automatic people recognition, has received increasing attention in the
recent years. Most of current analysis rely on the extraction  of
features characterizing the activity of single brain regions, like
power-spectrum estimates, thus neglecting possible temporal
dependencies between the generated EEG signals. However,
important physiological information can be extracted from the way
different brain regions are functionally coupled. In this study, we
propose a novel approach that fuses spectral coherence-based connectivity between different brain regions as a possibly viable biometric feature. The proposed approach is tested on a large dataset of subjects (N=108) during eyes-closed (EC) and
eyes-open (EO) resting state conditions. The obtained recognition
performances show that using brain connectivity leads to higher distinctiveness with respect to power-spectrum measurements, in both the experimental conditions.
Notably, a 100\% recognition accuracy is obtained in EC and EO when integrating functional connectivity between regions in the frontal lobe, while a lower 97.41\% is obtained in EC (96.26\% in EO) when fusing power spectrum information from centro-parietal regions.
Taken together, these results suggest that functional
connectivity patterns represent effective features for improving EEG-based biometric systems.
\end{abstract}


\section{Introduction}

\IEEEPARstart{E}{lectroencephalography} (EEG) signals provide relevant information
about individual differences related to brain anatomical and
functional traits as already pointed out in early neurophysiologic
studies \cite{Berkhout68}, \cite{VanDis79}. Although some isolated
attempts to discriminate people from their electrical brain activity have been performed in
the past \cite{Stassen80} only recently the scientific community has
started a more systematic investigation on the use of EEG signals as
human distinctive traits which can be potentially used in a
biometric system \cite{CampisiCOMPUTER12}. A variety of EEG
elicitation protocols for the purpose of automatic user recognition
have been implemented, ranging from resting states with eyes open
(EO) and eyes closed (EC) to self-paced mental imagery \cite{Marcel06}.

In particular, resting states are advantageous conditions for biometric applications since they don't require active involvement of the subject during the EEG recording, thus reducing inconvenience, fatigue and artifact occurrences. From a neurophysiological perspective, ongoing EEG activity during resting states elicits patterns of synchronous oscillations in specific frequency ranges (from 1 to 40 Hz) that share and support basic cognitive functions \cite{mantini2007}. Furthermore it has been suggested that EEG activity during resting wakefulness carries genetic information \cite{Vogel1979c} and personality correlates \cite{tomarken1992}.

The number of studies investigating EEG activity during resting states as a potential biometric marker has increased, though the obtained results suggest that more efforts should be done to improve its efficacy in terms of correct recognition performance (CRR) and/or sample size, i.e. the number of subjects to recognize. A comprehensive survey of the different methods, protocols and achieved results can be found in \cite{CampisiTIFS2014}.

As evidenced by this survey, the large part of the studies exploring EEG for biometric
purposes has focused on the extraction of features from single electrodes. However, complementary information can be obtained from the
temporal dependence between activities of different brain areas
\cite{Varela2001}. From a neuroscience perspective two brain regions
that exhibit coherent or correlated activities are supposed to
exchange information \cite{akam2010}, \cite{astolfi2004} and this
kind of communication represents nowadays a key element to understand the brain organization \cite{he2011}.
Many analytical tools are available to measure  statistical interdependence between brain signals that are based on different mathematical principles (e.g. correlation, information theory, phase coherence, Granger-causality), implemented in the time and frequency domain, capturing either linear or nonlinear changes \cite{david2004}, \cite{billinger2013}. These tools allow the estimation of the so-called functional brain connectivity \cite{friston2011}.

Interestingly enough, we know that the way
brain regions are functionally connected is not homogeneous, and
that specific connectivity patterns emerge during wakeful resting
state conditions \cite{laufs2003}. Our hypothesis is
that functional connectivity between EEG sensors could be a more
robust feature for biometric recognition purpose, compared to EEG
activity of channels considered separately. 
A previous study has tested, amongst others, time-domain connectivity measures between two frontal electrodes (Fp1 and Fp2) for the identification of 51 subjects \cite{Riera08}. The recognition accuracy, obtained separately with different classifiers, reached a maximum of CRR=31\% for cross-correlation and CRR=24\% for mutual information. In this regard, the large part of the studies using EEG for biometric purposes showed that the extraction of features describing the spectral content of the signals could give better recognition performance \cite{CampisiTIFS2014}. Hence, frequency-domain connectivity measures could be more appropriate for detecting those putative distinctive features.

Changes of EEG amplitudes in the same subject and during the same condition can occur due to physiologic circadian rhythms
\cite{aeschbach1999}, substance assumption \cite{sokhadze2008}, or
different recording technical solutions \cite{ferree2001}. In
contrast to univariate measures, such as power spectrum estimates, spectral coherence is a simple bivariate connectivity method that is not sensitive to the amplitude changes of the EEG oscillations. In fact, two signals may have different amplitudes and/or phases, but high
coherence occurs when this phase difference tends to remain constant
\cite{nunez2006}. Therefore, this property could play a critical
role in increasing the overall classification performance in
presence of large intra-subject EEG variability due to scale
factors. 

Another common issue in EEG-based biometric systems is the general tendency to
perform classification from single elements, e.g. power spectra and/or functional connectivity \cite{Poulos99b}, \cite{Su12},
\cite{DeVicoFallani2011},\cite{Riera08}. Less often the features space has been
enriched by including information from a restricted number of
a-priori selected elements \cite{CampisiWIFS11},
\cite{LaRoccaBIOSIG2013}. These parsimonious procedures appear certainly justified by
possible implications on industrial research for smart EEG systems
with few electrodes. However, such technical solution restricts the
features space and consequently can affect the overall classification
performance. Alternative methodological solutions integrating
information from multiple elements can be beneficial with the aim of
catching more robust and distinctive brain patterns, thus maximizing
the recognition rates. 

In this study we propose a fusion approach \cite{ross2006} to integrate at the match-score level the information obtained from the estimation of spectral coherence estimates.

The combined use of EEG spectral coherence and classification algorithms has been previously exploited in neuroscience to distinguish between healthy and diseased populations \cite{duffy2012}  or to determine changes between baseline and motor/cognitive tasks \cite{krusienski2012}. To the best of our knowledge, this is the first time that such a combined approach is also used for biometric purposes.


\section{Methods}
\label{subsec:methods}
\subsection{Dataset and Preprocessing}
\label{subsec:preprocessing} Scalp EEG signals were gathered from
the freely online database PhysioNet BCI \cite{PhysioNetBCI}. The
database consists of \(N=108\) healthy subjects recorded in two
different baseline conditions, i.e. 1-minute EO resting state and 1-minute EC resting state. In each condition,
subjects were comfortably seated on a reclining chair in a dimly lit
room. During EO they were asked to avoid ocular blinks in order to reduce
signal contamination. The EEG data were recorded with a 64-channel
system (BCI2000 system \cite{BCI2000}) with an original sampling
rate of \(160 \Hz\). All the EEG signals are here referenced to the
mean signal gathered from electrodes on the ear lobes. Data are
subsequently downsampled to \(100 \Hz\) after applying a proper anti-aliasing
low-pass filter to restrict the available frequency range up to \(50
\Hz\). The electrode positions on the scalp follow the standard
10-10 montage. 5 electrodes are excluded and only \(N_{ch}=56\)
electrodes (see Figure \ref{fig:PerfFusion}) are retained for the
subsequent analysis. These electrodes were selected because they
constitute a montage common to different available datasets that we
could use in future analysis. 

For each subject and condition
(EO, EC) the obtained EEG signals are segmented into \(N_T=6\)
consecutive non-overlapping epochs of 10 seconds. These epochs are
considered as different observations of the same mental state and
they are used to extract specific spectral features for the
assessment of person recognition. In particular, we consider the two
following methods, power spectral density and functional connectivity detailed in Sections \ref{subsec:PSD} and \ref{subsec:COH}, respectively.

\subsection{Power spectral density}\label{subsec:PSD}
Although parametric linear or non-linear EEG signal processing has
been recently investigated for biometric purposes
\cite{LaRoccaBIOSIG12}, \cite{poulos2002}, a non-parametric Fourier
Transform-based spectral analysis is chosen in this study due to its
obvious physical interpretation in terms of EEG rhythms.
Specifically the power spectral density (PSD) of the EEG signals was
extracted from each segmented epoch (\(10 s\)) by computing the
Welch's averaged modified periodogram. A sliding Hanning window of
\(1 s\), with an overlap of \(0.5 s\), is applied to improve the
estimation quality. The number of FFT points is set to \(100\) in
order to have a PSD estimate with a frequency resolution of \(1
\Hz\) (the frequency sampling is \(100 \Hz\)). The resulting PSD for
the electrode \(i\), with \(i=1,\cdots,N_{ch}\), is a feature vector of
\(N_{FT}=41\) elements characterizing the power of the EEG
oscillations from \(0\) up to \(40 \Hz\). In the present study we
consider a restricted range of frequency, namely \(1-40 \Hz\). This
choice covers the standard spectrum of physiologic EEG oscillations
from low (Delta \(1-3 \Hz\), Theta \(4-7 \Hz\)) to intermediate
(Alpha \(8-14 \Hz\)) and high frequency bands (Beta
\(15-29 \Hz\), Gamma \(30-40 \Hz\)). Each segmented epoch is
finally characterized by \(N_{E}^{(PSD)}=N_{ch}\) feature vectors
\(\hat\zeta^{(PSD)}\) of \(N_F=40\) elements, where \(N_{ch}=56\) is the
total number of electrodes.

\subsection{Functional connectivity}\label{subsec:COH}
In this study, functional connectivity is estimated by calculating
the spectral coherence (COH) \cite{andres1999}. This method is
frequently used due to its practical and intuitive interpretation.
Spectral coherence quantifies the level of synchrony between two
stationary signals at a specific frequency \(f\). 
Given two signals
obtained from channels \(i\) and \(j\), the spectral coherence
\(COH_{i,j}(f)\) for a particular frequency \(f\) is computed as
follows:
\begin{equation}
COH_{i,j}(f)=\frac{|S_{i,j}(f)|^2}{S_{i,i}(f)\cdot S_{j,j}(f)}.
\end{equation}
where \(S_{i,j}(f)\) is the cross-spectrum of the signals acquired from channels \(i\) and \(j\), while \(S_{i,i}(f)\) and \(S_{j,j}(f)\) are the respective autospectra.
 By definition \(COH_{i,j}(f)\) ranges between \(0\), which
corresponds to no synchrony at the frequency \(f\) and \(1\), which
corresponds to maximum synchrony at the frequency \(f\). Here,
\(S_{i,j}(f), S_{i,i}(f)\) and \(S_{j,j}(f)\) are computed by means
of the Welch's averaged modified periodogram, with same parameters
used for the computation of the PSD (see Section \ref{subsec:PSD}). 
In particular the use of 1s Hanning windows should improve the stationarity of the segmented EEG signals \cite{fingelkurts2004}.
For each electrode pair, we consider a feature vector
\(\hat\zeta^{(COH)}\) consisting of \(N_F=40\) \(COH\) values ranging
from \(f=1 \Hz\) to \(f=40 \Hz\). The total number of
electrodes \(N_{ch}\) being equal to \(56\), we have that \(i=1,\cdots,N_{ch}-1\) and
\(j=i+1,\cdots,N_{ch}\) and each segmented epoch (\(10 s\)) is finally
characterized by \(N_{E}^{(COH)}=\frac{N_{ch}\cdot(N_{ch}-1)}{2}=1540\)
features vectors of \(N_F\) elements.  
\subsection{Classifier}\label{subsec:Classifier}
We use a classification approach to predict the
class, namely the subject identity, to which the observed feature
vector \(\hat\zeta\) belongs. The model we use for the discriminant
analysis assumes that the feature vectors \(\hat\zeta\) form a
Gaussian mixture distribution. 

For this reason, before the classification, we apply a Fisher's Z transformation to the COH values in order to normalize their distributions \cite{amjad1997}. A logarithmic transformation is instead applied to the PSD values \cite{gasser1982}.

A Mahalanobis distance-based classifier is then used to perform identification. This method requires the computation of the covariance matrix of the feature vectors of each class. Given the few observations for each subject (6 epochs) the covariance matrices cannot be robustly computed. Therefore we use a common procedure consisting in the simplification to equal covariance matrices \cite{schurmann1996}. Specifically, a pooled covariance matrix is obtained by merging the class-specific distributions of the feature vectors after removing their mean value. Notably, the normalizing transformation that we apply supports such approximation.
A partition of the \(N_T\times N\) normalized feature vectors extracted from the dataset is used to enroll the subjects and generate templates representing the class distributions. 

A cross-validation framework is here considered to assess the
recognition performance. In each cross-validation run, for each
subject \(m\), \(5\) epochs are used to generate the class
distributions (i.e. enrollment phase), while the remaining  epoch is employed to perform the
identification (i.e. test phase). A number \(N_r=6\) of runs are provided, considering all possible partitions (leave-one-out framework).
The Mahalanobis distances are computed between each observation \(m\) in the
test dataset, and the class distributions \(n\) obtained in the
enrolling, according to the formula:
\begin{equation}
d_{m,n}=(\hat\zeta_m-\mu_n)\Sigma^{-1}(\hat\zeta_m-\mu_n)^T
\end{equation}
with \(m,n=1,\cdots,N\), \(N\) being the number of classes
(subjects), \(\hat\zeta_m\) the observed feature vector from subject
\(m\), \(\mu_n\) the mean vector for the class distribution \(n\),
and \(\Sigma\) the pooled covariance matrix. 

We use the misclassification (or confusion) matrix
\(\bold{M}\) to evaluate the recognition performance. Each column of this matrix \(N \times N\) represents the instances in a predicted subject identity (i.e. class), while each row represents the instances in an actual class. 
For a given subject \(m\), the predicted subject identity \(\hat{n}\) is obtained according to:
\begin{equation}
\hat{n} = \arg \min_{n} d_{m,n}.
\end{equation}

Eventually, for each run the correct recognition rate (CRR) is defined as the average over the diagonal of
the resulting misclassification matrix \(\mathbf{M}\):
\begin{equation}
CRR=\left(\frac{1}{N}\sum_{n=1}^{N}M[n,n]\right)\times100.
\end{equation}
\subsection{Match score fusion}\label{subsec:Fusion}
Performance in terms of correct recognition percentage is computed
as described in the previous Section for both the PSD based approach
and the connectivity estimate based method, separately. We can
observe that different elements, both channels and channel pairs,
show to be distinctive for different groups of
subjects correctly identified. Therefore we try to obtain complementary information
considering activities of different brain areas, known not to be
homogeneous, in order to improve overall performance. Through the
fusion of single-element information we obtain patterns of brain
activity, supposed to be a more robust characterization
representative of brain organization of specific subjects. In this
regard we perform a fusion at the match score level, considering
the sum of scores related to different elements:
\begin{equation}
\label{eq:sumrule}
S_{m,n}=\sum_{{e}\in E}\frac{1}{d_{m,n}^{e}}
\end{equation}
where \(E\) is a selection of elements from the set of \(56\)
channels for \(\hat\zeta^{(PSD)}\), or from \(1540\) channel pairs
when considering \(\hat\zeta^{(COH)}\). We then compute the
misclassification matrix \(\mathbf{M}\) to evaluate recognition accuracy 
as described in Section \ref{subsec:Classifier} by maximizing
\(S_{m,n}\) with respect to \(n\). We select the elements to
consider in the match score fusion according to a forward-backward
approach. 
Specifically only elements which improve accuracy are retained in the fusion. First, all the elements are sorted in a descending order of accuracy according to the single-element classification results. Starting from the first one, each single element is then added stepwise in the subset \(E\) to compute equation (\ref{eq:sumrule}). If the inclusion of the i-th element improves the resulting accuracy then it is retained for the final fusion, otherwise it is removed (see Fig. \ref{fig:PerfFusion}). More details about the pseudo-code implementation of the match score fusion can be found in the Supplementary File 1).

Each step of the fusion algorithm is computed within a leave-on-out
cross-validation framework, as discussed in Section
\ref{subsec:Classifier}. Specifically, for each tested subset \(E\) of
elements, related to a specific step of the fusion algorithm,
\(S_{m,n}\) is computed for each of the \(6\) partitions of the
dataset. Then the CRRs related to the different partitions are
averaged together to obtain the performance related to the particular step. If that performance represents an improvement compared to the previous step the related element is retained in \(E\) for the final fusion. 
This analysis is carried out separately for three cerebral zones,
namely frontal (F), central (C), and parieto-occipital (P), in order to compare
performance and distinctive activity patterns among them. These
zones are selected according to previous studies which
showed an improvement of recognition accuracy spanning the scalp
from the anterior part of the head to the posterior area
\cite{LaRoccaBIOSIG12}, \cite{DeVicoFallani2011}.

\section{Results and Discussion}\label{sec:Results}
In this section the results of the analysis performed for subject
recognition based on EEG signals are reported and discussed.
Resting-state conditions EO and EC are investigated separately and
the related outcome is here compared for all the performed tests.
Two different characterizations of the brain signals are considered
as distinctive features to test within the provided biometric
framework. In particular, after the preprocessing
described in Section \ref{subsec:preprocessing}, PSD and COH
estimates are obtained as reported in Sections \ref{subsec:PSD} and
\ref{subsec:COH}. As previously pointed out, PSD measures the activity of single brain regions while COH rather measures their functional connectivity. 
Then, two different sets of feature vectors,
\(\hat\zeta^{(PSD)}\) and \(\hat\zeta^{(COH)}\), are extracted
according to the previously described methods. A classification based on Mahalanobis distance, detailed in Section
\ref{subsec:Classifier}, is then carried out to evaluate the
distinctiveness of the two considered feature vectors in all tested
conditions, in terms of CRR.
\begin{figure}[t!]
\centering
\includegraphics[width=90.0mm]{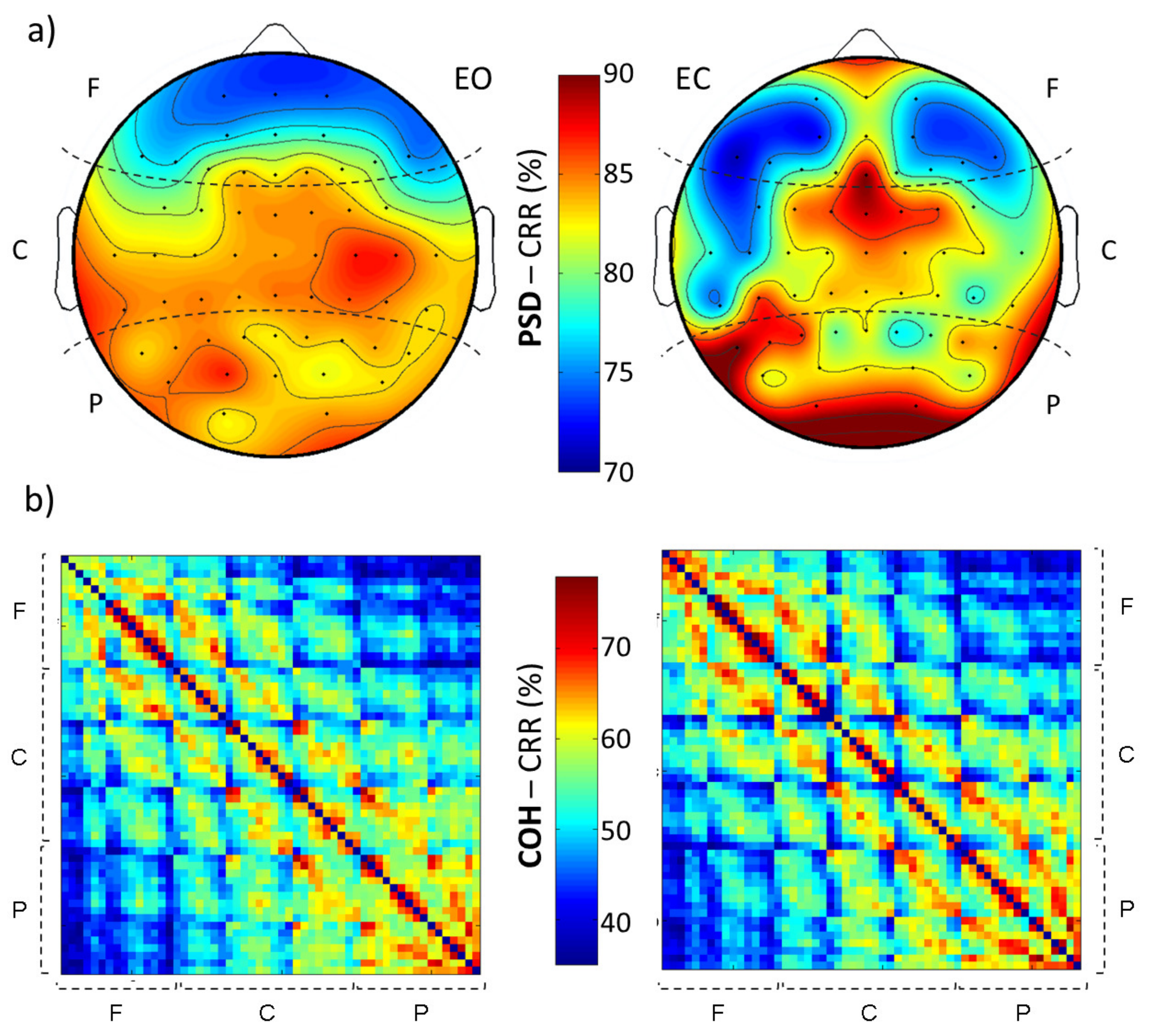}
    \caption{Spatial distribution of CRR values obtained considering
    a) PSD features from single EEG channel and b) COH features from single channel pairs. In b) the elements of the adjacency matrices code all the possible EEG channel pairs. They are organized in order to highlight the connectivity within and between three zones: frontal (F), central (C), and parieto-occipital (P). The two analyzed conditions EO (on the left) and EC (on the right) are reported for each of the two investigated spectral features.}
    \label{fig:PerfSingle}
\end{figure}
\subsection{Single-element classification}
A preliminary test on distinctiveness related to each feature is
reported in Fig. \ref{fig:PerfSingle}. The CRR values obtained
within a cross-validation framework, as described in Section
\ref{subsec:Classifier}, are shown in false colors. In particular
the scalp maps shown in Fig. \ref{fig:PerfSingle} (a) represent the spatial distribution of the CRR values obtained through single channel PSD
features, in the EO and EC conditions. The adjacency matrices shown in Fig. \ref{fig:PerfSingle} (b) report the CRR values of the COH features for each
channel pair. Given the \(N_{ch}\times N_{ch}\) adjacency matrix
\(\mathbf{Adj}\), each element \(Adj_{i,j}\) represents the CRR
obtained considering the COH between channels \(i,j\) as feature
vector. According to this representation, the position of the EEG
channels over the head is coded by the \(x\) and \(y\) axes of
the adjacency matrix.
The CRR scalp maps for the PSD features show that the most distinctive regions appear in
the central part of the head (C) during EO (max CRR=86.91\%),while
the parieto-occipital zone (P) is more predominant during the EC
condition (max CRR=90.49\%). The better distinctiveness of the
posterior areas during EC condition was already reported in
previous EEG studies \cite{DeVicoFallani2011}, \cite{CampisiWIFS11}, \cite{LaRoccaBIOSIG12}
and it is probably related to well-known physiological increase of
the parieto-occipital Alpha power in such condition \cite{niedermeyer2005chp9}. A possible explication for this
evidence is that eyes-closed resting states interrupt the visual
processing while enhancing endogenous and autonomic related brain
activity \cite{Barry}, which reflects the influence of genetic
factors \cite{bodenmann2009}.
For the CRR adjacency matrices in
the EC condition the best discriminant channel pairs are
mainly located in the parieto-occipital zone P (max CRR=78.5\%). In
the EO condition, the CRR values are generally lower compared to EC
(max CRR=75.86\%). In both conditions, short-range functional
connectivity carries more distinctive information as can be observed
by the tendency of the highest CRR values to stay close to the main
diagonal of the \(\mathbf{Adj}\) matrices. 

\subsection{Match-score fusion}
\begin{figure}[t!]
\centering
\includegraphics[width=65.0mm]{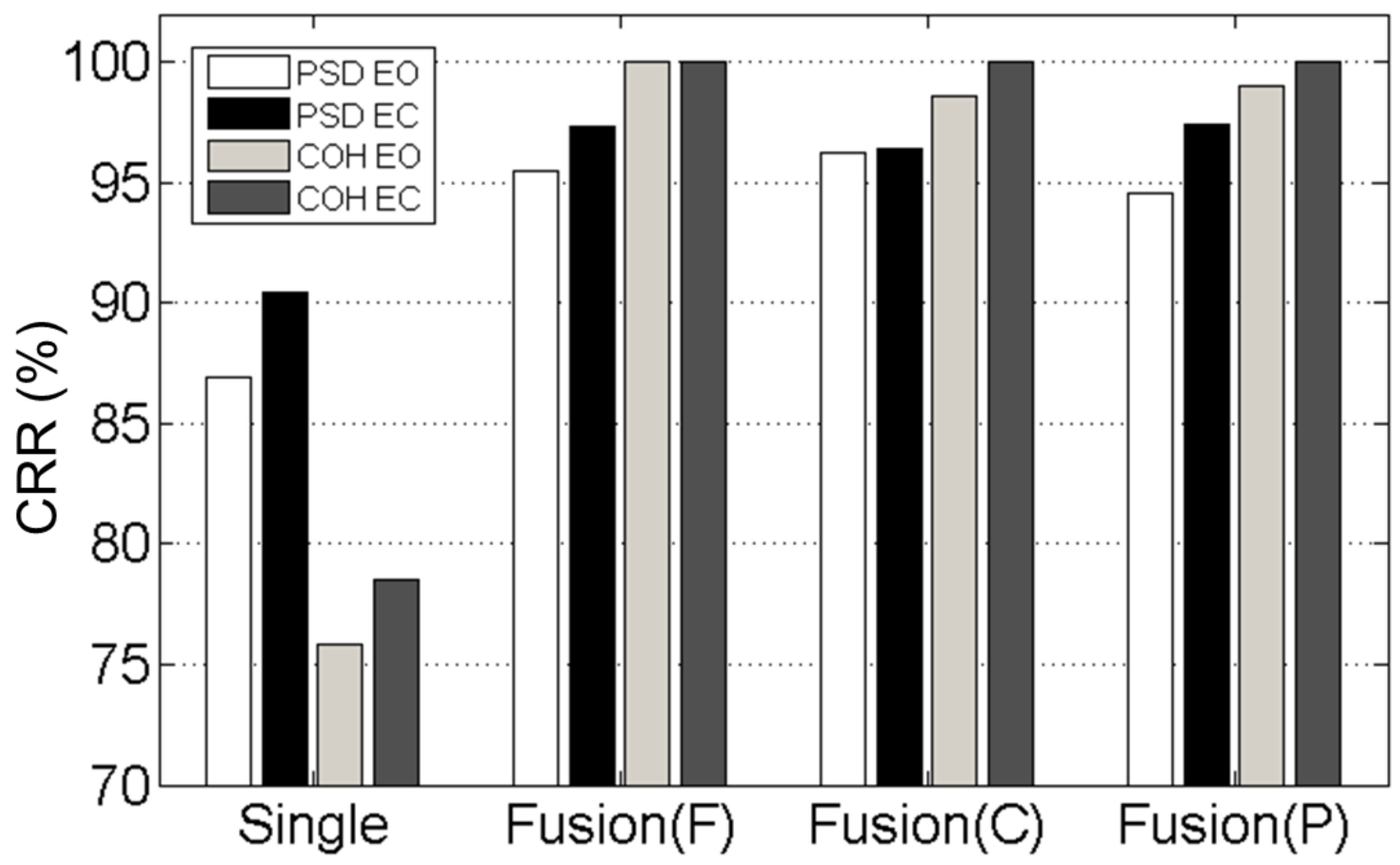}
    \caption{Performances in terms of CRR (y-axis) obtained considering single-element classification versus match-score fusion in each cerebral zone (x-axis). The color of the bars codes the spectral feature and the condition according to the legend.}  
\label{fig:Perf}
\end{figure}

To improve performance, a fusion of the elements at the match score
level is obtained for each cerebral zone (F, C and P), within the
same cross-validation framework described in Section \ref{subsec:Fusion}.
The related improvements with respect to the single element approach
are shown in Fig. \ref{fig:Perf}. Here the two conditions EO and EC, and the two considered features PSD and COH,
are compared. In general, a dramatic improvement with respect to the single element approach is obtained
for each zone, condition and feature through the match score
fusion approach. 
This is particularly true for the COH features, for which the fusion allows a perfect recognition rate \(CRR=100\%\) for the EC condition (in all the zones) and for the EO condition (in the frontal zone).

\begin{figure}[t!]
\centering
\includegraphics[width=90.0mm]{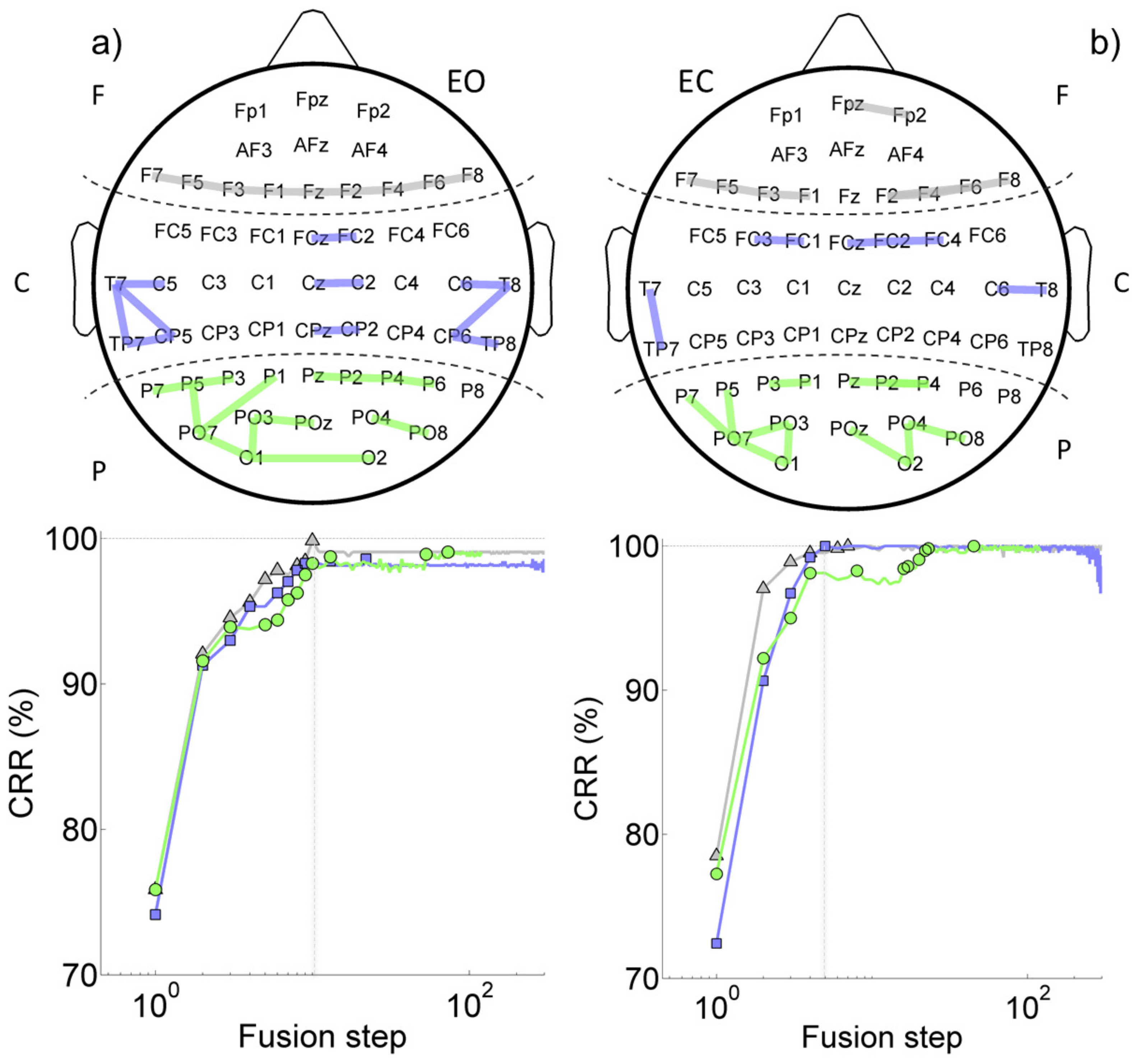}
    \caption{Distinctive functional connectivity patterns in each cerebral zone (on the top) and related steps for the match score fusion selection (on the bottom). Results for the EO and EC to conditions are reported in a) and b) respectively. The color of each line in the bottom panels codes the CRR values obtained for different cerebral zones, according to the top panels. Symbol markers highlight the fusion steps that increased the overall CRR accuracy. Different symbols are used for different cerebral zones (F-triangles, C-circles, P-squares). X-axes are put into logarithmic scales for the sake of representation. }
    \label{fig:PerfFusion}
\end{figure}

Fig. \ref{fig:PerfFusion} shows the
optimal combination of channel pairs in every zone for COH features (see Section
\ref{subsec:Fusion}). Results are shown for the two conditions EO (a) and EC (b). 
The plots shown on the bottom part of the figure represent the steps
of the match score fusion. Here, each highlighted symbol represents a
subsequent improvement of the overall CRR, which leads to include
the related channel pair in the final distinctive
connectivity pattern represented in the upper part of the figure. It
can be noticed that the maximum value of CRR is achieved more
rapidly for the EC condition, as shown by the vertical lines in the plots of the bottom part of the figure. The resulting distinctive connectivity patterns consist mainly of short-range COH elements,
a result which is in line with the previous outcome reported in the
single-element classification analysis. 
The topology of the more discriminant COH elements reveals a hemispheric
symmetry with respect to the longitudinal line. In particular, the predominance of the frontal electrode pairs (line \(F7-F8\)) can be observed in both the EO and EC conditions. Neurophysiological evidence shows that specific genetic factors can influence EEG frontal activity \cite{zietsch2007}. Thus the highly
discriminant spectral coherence observed in this zone could in part
reflect a subjective distinctiveness of brain functioning. 
We also report the involvement of temporo-parietal and dorsal
centro-parietal electrodes of the central zone C. Brain regions near the
temporo-parietal junction play a specific role in self-other
distinction processes and in representing thoughts, beliefs,
desires, and emotions \cite{saxe2003}, influenced by a combination
of biological and environmental factors. 
A comprehensive analysis reporting the fusion steps for the spectral coherence features in all the conditions and zones is
detailed in Table \ref{tab:COH}, along with a
description of the obtained identification accuracy and topology. A predominance of short-range connectivity
characterizes the distinctive patterns for all the considered brain zones. 

Notably, the inclusion of the long-range (i.e. inter-zone) connectivity in the fusion algorithm does not improve the recognition performance  (Supplementary File 2).

The superiority of short-range over long-range connectivity could be partially imputed to volume condition effects, which are known to affect spectral coherence measurements \cite{nolte2004}. Although removing those effects is important in general to estimate the real interaction between the cortical generators, this could not represent an issue in our study. In fact, volume conduction effects depend on the morphology and electrical conductibility of the subject's head structures. In this regard, any possible volume conduction contributions on the EEG signals could instead represent a relevant personal trait to be exploited for the biometric recognition. As a partial confirmation of our claim we reported that the recognition performance of the imaginary coherence (robust to volume conduction effects \cite{nolte2004}) is significant lower than standard spectral coherence (Supplementary File 3). 


To sum up, the results of the herein proposed analysis
show that a perfect identification of 108 subjects (\(CRR=100\%\)) can be obtained considering spectral coherence features within specific
regions of the head, and fusing the respective information at the
match score level. This result outperforms the state-of-the-art recognition performance obtained with EEG during resting states, notably a \(CRR=98.73\%\) for a dataset of 45 subjects \cite{LaRoccaBIOSIG12} and a \(CRR=97.5\%\) for a dataset of 40 subjects \cite{Su12}.

\subsection{Limitations and possible solution}

The proposed approach presents methodological and technical limitations that should be taken into account in scenarios different from the presented experimental protocol. First, the spectral coherence requires EEG signals to be (quasi)stationary.  Appropriate short time windows should be selected by testing the stationarity of the signals \cite{kwiatkowski1992}. Alternatively different connectivity methods that don't require stationarity (e.g. wavelet-based) can be used \cite{le_van_quyen2001}. Second, the Mahalanobis distance-based classifier assumes that features are Gaussian distributed. Possible deviation from Gaussianity should be then compensated by applying appropriate data transformations \cite{amjad1997} or using reduced polynomial regression-based classifiers (PRC) \cite{scarano2012}. Third, the fusion approach implies that many EEG sensors need to be placed on the scalp. This affects the design of the biometric system  and the time needed to establish a good skin-sensors electrical contact. Possible  solutions can come from the technology development related to dry and miniaturized sensor helmets, or non-contact biosensors \cite{chi2010}. Finally, most of the computational time is spent during the off-line enrollment and definition of the fusion steps (around 20 min on a standard personal computer). Though it is not the aim of this study, we envisage that possible optimizations can be obtained by parallelizing the single-element ranking and the cross-validation runs.

\begin{center}
 \begin{table*}[t!]
 {\scriptsize
\hspace{3.25cm}
 \begin{tabular}{|c|@{}l@{}|@{}l@{}|@{}l@{}|@{}l@{}|@{}l@{}|@{}l@{}|@{}l@{}|@{}l@{}|@{}l@{}|@{}l@{}|@{}l@{}|@{}l@{}|}
 \hline
 \multicolumn{13}{|c|}{\bf{\small{Eyes-Open}}}\\ \hline
\bf{F}     &72,74  &91,43 &94,55  &97,2  &97,82  &99,06  &99,84  &100  &   &   &  & \\
      &F2-F4  &F1-F3 &F4-F6 &F3-F5 &F6-F8 &F1-Fz  &F5-F7 &F2-Fz &  &  & &     \\
\hline
\bf{C}     &74.14  &91.28 &92.99   &95.33  &96.26  &97.04  &97.82  &98.29  &98.44   &98.6   &  & \\
     &TP7-T7 &CP5-TP7 &C5-T7  &T8-C6 &FC2-FCz &C2-Cz &CP2-CPz&CP5-T7 &TP8-CP6 &CP6-T8  & &\\
\hline
\bf{P}    &75.86  &91.59 &93.92  &94.08  &94.39  &95.79  &96.26  &97.51  &98.29   &98.75   &98.91  &99.07 \\
     &P7-P5  &PO4-PO8&O1-PO3&O1-PO7 &P2-P4  &PO7-P5  &O1-O2 &P2-Pz  &P3-P5   &P4-P6   &POz-PO3&PO7-P1 \\
\hline \hline \multicolumn{13}{|c|}{\bf{\small{Eyes-Closed}}}\\
\hline
\bf{F}    &77.26  &97.04  &97.82 &98.6  &98.75  &99.06  &99.69 &100  &   &   &  & \\
     &F2-F4  &F3-F5  &F1-F3 &F5-F7  &F6-F8  &F4-F6  &Fp2-Fpz  &F2-F6  & &  &  &\\
\hline
\bf{C}    &72.43  &90.65  &96.73 &99.22  &100  &  &  &  &   &   &  & \\
     &TP7-T7 &FC2-FCz&FC2-FC4&C6-T8  &FC1-FC3 &  && &  &  &&\\
\hline
\bf{P}   &78.5  &92.21  &95.01 &98.13  &98.29  &98.44  &98.6  &99.06  &99.69   &99.84   &100  & \\
     &PO3-O1 &PO4-PO8 &PO7-O1&PO3-PO7 &P2-P4  &PO7-P7 &PO7-P5&P1-P3  &O2-PO4  &O2-POz   &P4-Pz &\\
\hline
\end{tabular}
 \vspace{.3cm}
 \caption{CRRs for COH features, obtained through the fusion of the channel pairs reported in each row. Each column represents a step of the fusion, and the related accuracy achieved is reported together with the channel pair considered in that step (see Section \ref{subsec:Fusion}). Results for EO, EC, and the three investigated brain zones are shown.}
 \label{tab:COH}
 }
 \end{table*}
 \end{center}

\section{Conclusions}
\label{sec:Conlusions} In the past few years there has been a
growing interest in EEG-based biometric systems. Compared to other traits that
are usually considered in biometrics (e.g. fingerprints, iris,
voice, signatures, etc.), EEG activity presents two main advantages,
among others: i) it's harder to steal  and ii) it's a dynamic measure, thus
allowing for constant recognition and mental state monitoring
\cite{CampisiCOMPUTER12}. Despite such interest, classification
performance decreases from maximum accuracy when the number of
people to recognize becomes high, i.e. \(N>100\) \cite{Brigham10},
\cite{Palaniappan07}. Possible causes rely on the methods that are
commonly used to extract characteristic features from the EEG
signals. Indeed, the majority of these methods only consider the
activity of a single brain region without taking into account its
dynamic relationship with other regions. It is worth pointing out
that the human brain is known to be an interconnected system where
different specialized areas continuously exchange information
through stable synchronous connectivity patterns \cite{Varela2001}.
In the present study we propose to exploit such synchronous
''communication'' between different EEG sensors, under the
hypothesis that this information could exhibit stronger intra-class
invariant properties. Specifically,  spectral coherence estimates within specific zones (i.e. frontal, central,
parieto-occipital) are integrated according to a fusion at match
score level \cite{ross2006} based on the sum rule. 

We test the proposed
approach on a large number of subjects (N=108) having been recorded
with a high-density EEG system with 56 electrodes available. Taken
together, the obtained results indicate that the combined use of
spectral coherence and fusion algorithms significantly improves the
overall performance compared to existing standard techniques. 
Notably a perfect recognition can be achieved in both eyes-closed and eyes-open resting states by considering only 15 EEG sensors in the frontal region of the head. This also represents a technical advantage for future implementation of smart EEG-based biometric helmets.

Although connectivity-based approaches have not received much attention in this field, we
suggest that they will probably turn out to be effective invariant
features to further develop robust EEG-based user-recognition systems.

\bibliographystyle{IEEEtran}
{\small
\bibliography{IEEEBrain}}

\begin{thebibliography}{10}
\providecommand{\url}[1]{#1}
\csname url@samestyle\endcsname
\providecommand{\newblock}{\relax}
\providecommand{\bibinfo}[2]{#2}
\providecommand{\BIBentrySTDinterwordspacing}{\spaceskip=0pt\relax}
\providecommand{\BIBentryALTinterwordstretchfactor}{4}
\providecommand{\BIBentryALTinterwordspacing}{\spaceskip=\fontdimen2\font plus
\BIBentryALTinterwordstretchfactor\fontdimen3\font minus
  \fontdimen4\font\relax}
\providecommand{\BIBforeignlanguage}[2]{{%
\expandafter\ifx\csname l@#1\endcsname\relax
\typeout{** WARNING: IEEEtran.bst: No hyphenation pattern has been}%
\typeout{** loaded for the language `#1'. Using the pattern for}%
\typeout{** the default language instead.}%
\else
\language=\csname l@#1\endcsname
\fi
#2}}
\providecommand{\BIBdecl}{\relax}
\BIBdecl

\bibitem{Berkhout68}
J.~Berkhout and D.~O. Walter, ``Temporal stability and individual differences
  in the human {EEG}: An analysis of variance of spectral values,'' \emph{IEEE
  Trans. Biomed. Eng.}, vol.~15, no.~3, pp. 165--168, 1968.

\bibitem{VanDis79}
H.~Van~Dis, M.~Corner, R.~Dapper, G.~Hanewald, and K.~H, ``Individual
  differences in the human electroencephalogram during quiet wakefulness,''
  \emph{Electroenceph. and clin. neurophys.}, vol.~47, pp. 87--94, 1979.

\bibitem{Stassen80}
H.~H. Stassen, ``Computerized recognition of persons by {EEG} spectral
  patterns,'' \emph{Electroenceph. and clin. neurophys.}, vol.~49, no.~1, pp.
  190--194, 1980.

\bibitem{CampisiCOMPUTER12}
P.~Campisi, D.~La~Rocca, and G.~Scarano, ``{EEG} for automatic person
  recognition,'' \emph{IEEE Computer}, vol.~45, no.~7, pp. 87--89, 2012.

\bibitem{Marcel06}
S.~Marcel and J.~del R.~Millan, ``Person authentication using brainwaves
  ({EEG}) and maximum a posteriori model adaptation,'' \emph{IEEE Trans.
  Pattern Anal. Mach. Intell.}, vol.~29, no.~4, pp. 743--748, 2006.

\bibitem{mantini2007}
D.~Mantini, M.~G. Perrucci, C.~D. Gratta, G.~L. Romani, and M.~Corbetta,
  ``Electrophysiological signatures of resting state networks in the human
  brain,'' \emph{Proc. Natl. Acad. Sci. U.S.A.}, vol. 104, no.~32, pp.
  13\,170--13\,175, Aug 2007.

\bibitem{Vogel1979c}
F.~Vogel and E.~Schalt, ``The electroencephalogram ({EEG}) as a research tool
  in human behavior genetics: Psychological examinations in healthy males with
  various inherited {EEG} variants,'' \emph{Hum. Genet.}, vol.~47, pp. 81--111,
  1979.

\bibitem{tomarken1992}
A.~J. Tomarken, R.~J. Davidson, R.~E. Wheeler, and L.~Kinney, ``Psychometric
  properties of resting anterior {EEG} asymmetry: Temporal stability and
  internal consistency,'' \emph{Psychophysiology}, vol.~29, no.~5, pp.
  576--592, 1992.

\bibitem{CampisiTIFS2014}
P.~Campisi and D.~La~Rocca, ``Brain waves for automatic biometric based user
  recognition,'' \emph{{IEEE} Trans. Inf. Forensics Security}, 2014, in press.

\bibitem{Varela2001}
F.~Varela, J.-P. Lachaux, E.~Rodriguez, and J.~Martinerie, ``The brainweb:
  Phase synchronization and large-scale integration,'' \emph{Nat. Rev.
  Neurosci.}, vol.~2, no.~4, pp. 229--239, Apr. 2001.

\bibitem{akam2010}
T.~Akam and D.~M. Kullmann, ``Oscillations and filtering networks support
  flexible routing of information,'' \emph{Neuron}, vol.~67, no.~2, pp.
  308--320, 2010.

\bibitem{astolfi2004}
L.~Astolfi, F.~Cincotti, D.~Mattia, S.~Salinari, C.~Babiloni, A.~Basilisco,
  P.~M. Rossini, L.~Ding, Y.~Ni, B.~He \emph{et~al.}, ``Estimation of the
  effective and functional human cortical connectivity with structural equation
  modeling and directed transfer function applied to high-resolution eeg,''
  \emph{J. Magn. Reson. Imaging}, vol.~22, no.~10, pp. 1457--1470, 2004.

\bibitem{he2011}
B.~He, L.~Yang, C.~Wilke, and H.~Yuan, ``Electrophysiological imaging of brain
  activity and connectivity—challenges and opportunities,'' \emph{IEEE Trans.
  Biomed. Eng.}, vol.~58, no.~7, pp. 1918--1931, 2011.

\bibitem{david2004}
O.~David, D.~Cosmelli, and K.~J. Friston, ``Evaluation of different measures of
  functional connectivity using a neural mass model,'' \emph{Neuroimage},
  vol.~21, no.~2, pp. 659--673, 2004.

\bibitem{billinger2013}
M.~Billinger, C.~Brunner, and G.~R. Müller-Putz,
  ``\BIBforeignlanguage{eng}{Single-trial connectivity estimation for
  classification of motor imagery data},''
  \emph{\BIBforeignlanguage{eng}{Journal of neural engineering}}, vol.~10,
  no.~4, p. 046006, Aug. 2013, {PMID:} 23751454.

\bibitem{friston2011}
K.~J. Friston, ``\BIBforeignlanguage{eng}{Functional and effective
  connectivity: a review},'' \emph{\BIBforeignlanguage{eng}{Brain connect.}},
  vol.~1, no.~1, pp. 13--36, 2011.

\bibitem{laufs2003}
H.~Laufs, K.~Krakow, P.~Sterzer, E.~Eger, A.~Beyerle, A.~Salek-Haddadi, and
  A.~Kleinschmidt, ``Electroencephalographic signatures of attentional and
  cognitive default modes in spontaneous brain activity fluctuations at rest,''
  \emph{Proc. Natl. Acad. Sci. U.S.A.}, vol. 100, no.~19, pp. 11\,053--11\,058,
  2003.

\bibitem{Riera08}
A.~Riera, A.~Soria-Frisch, M.~Caparrini, C.~Grau, and G.~Ruffini, ``Unobtrusive
  biometric system based on electroencephalogram analysis,'' \emph{EURASIP J.
  Adv. Signal Process}, 2008.

\bibitem{aeschbach1999}
D.~Aeschbach, J.~R. Matthews, T.~T. Postolache, M.~A. Jackson, H.~A. Giesen,
  and T.~A. Wehr, ``Two circadian rhythms in the human electroencephalogram
  during wakefulness,'' \emph{Am. J. Physiol. Regul. Integr. Comp. Physiol.},
  vol. 277, no.~6, pp. R1771--R1779, 1999.

\bibitem{sokhadze2008}
T.~M. Sokhadze, R.~L. Cannon, and D.~L. Trudeau, ``Eeg biofeedback as a
  treatment for substance use disorders: review, rating of efficacy and
  recommendations for further research,'' \emph{J. Neurother.}, vol.~12, no.~1,
  pp. 5--43, 2008.

\bibitem{ferree2001}
T.~C. Ferree, P.~Luu, G.~S. Russell, and D.~M. Tucker, ``Scalp electrode
  impedance, infection risk, and eeg data quality,'' \emph{Clin.
  Neurophysiol.}, vol. 112, no.~3, pp. 536--544, 2001.

\bibitem{nunez2006}
P.~L. Nunez, \emph{Electric fields of the brain: the neurophysics of
  EEG}.\hskip 1em plus 0.5em minus 0.4em\relax Oxford University Press, 2006.

\bibitem{Poulos99b}
M.~Poulos, M.~Rangoussi, and N.~Alexandris, ``Neural network based person
  identification using {EEG} features,'' in \emph{Proc. IEEE Int. Conf. Acoust.
  Speech Signal Process.}, vol.~2, 1999, pp. 1117--1120.

\bibitem{Su12}
F.~Su, H.~Zhou, Z.~Feng, and J.~Ma, ``A biometric-based covert warning system
  using {EEG},'' in \emph{Proc. 5th IAPR Int. Conf. on Biometrics (ICB)}, 2012,
  pp. 342--347.

\bibitem{DeVicoFallani2011}
F.~De~Vico~Fallani, G.~Vecchiato, J.~Toppi, L.~Astolfi, and F.~Babiloni,
  ``Subject identification through standard eeg signals during resting
  states,'' in \emph{Proc. IEEE Int. Conf. Eng. Med. Biol. Soc.}, 2011, pp.
  2331--2333.

\bibitem{CampisiWIFS11}
P.~Campisi, G.~Scarano, F.~Babiloni, F.~De~Vico~Fallani, S.~Colonnese,
  E.~Maiorana, and F.~L., ``Brain waves based user recognition using the "eyes
  closed resting conditions" protocol,'' in \emph{IEEE WIFS}, Nov. 2011, pp.
  1--6.

\bibitem{LaRoccaBIOSIG2013}
D.~La~Rocca, P.~Campisi, and J.~Sole-Casals, ``Eeg based user recognition using
  bump modelling,'' in \emph{Proc. Int. Conf. BIOSIG}, 2013, pp. 1--12.

\bibitem{ross2006}
A.~A. Ross, K.~Nandakumar, and A.~K. Jain, \emph{Handbook of
  multibiometrics}.\hskip 1em plus 0.5em minus 0.4em\relax Springer, 2006.

\bibitem{duffy2012}
F.~H. Duffy and H.~Als, ``\BIBforeignlanguage{eng}{A stable pattern of {EEG}
  spectral coherence distinguishes children with autism from neuro-typical
  controls - a large case control study},''
  \emph{\BIBforeignlanguage{eng}{{BMC} Med.}}, vol.~10, no.~1, p.~64, 2012.

\bibitem{krusienski2012}
D.~J. Krusienski, D.~J. {McFarland}, and J.~R. Wolpaw, ``Value of amplitude,
  phase, and coherence features for a sensorimotor rhythm-based brain-computer
  interface,'' \emph{Brain Res. Bull.}, vol.~87, no.~1, pp. 130--134, Jan.
  2012.

\bibitem{PhysioNetBCI}
\BIBentryALTinterwordspacing
Database physionet bci. [Online]. Available:
  \url{http://www.physionet.org/pn4/eegmmidb/}
\BIBentrySTDinterwordspacing

\bibitem{BCI2000}
\BIBentryALTinterwordspacing
Bci2000 system. [Online]. Available: \url{http://www.bci2000.org}
\BIBentrySTDinterwordspacing

\bibitem{LaRoccaBIOSIG12}
D.~La~Rocca, P.~Campisi, and G.~Scarano, ``{EEG} biometrics for individual
  recognition in resting state with closed eyes,'' in \emph{Proc. Int. Conf.
  BIOSIG}, 2012, pp. 1--12.

\bibitem{poulos2002}
M.~Poulos, M.~Rangoussi, N.~Alexandris, and A.~Evangelou, ``Person
  identification from the eeg using nonlinear signal classification,''
  \emph{Methods Inf Med.}, vol.~41, no.~1, pp. 64--75, 2002.

\bibitem{andres1999}
F.~G. Andres and C.~Gerloff, ``Coherence of sequential movements and motor
  learning,'' \emph{J. Clin. Neurophysiol.}, vol.~16, no.~6, p. 520, 1999.

\bibitem{fingelkurts2004}
A.~A. Fingelkurts and A.~A. Fingelkurts, ``Making complexity simpler:
  multivariability and metastability in the brain,'' \emph{Int. J. Neurosci.},
  vol. 114, no.~7, pp. 843--862, 2004.

\bibitem{amjad1997}
A.~Amjad, D.~Halliday, J.~Rosenberg, and B.~Conway, ``An extended difference of
  coherence test for comparing and combining several independent coherence
  estimates: theory and application to the study of motor units and
  physiological tremor,'' \emph{J. Neurosci. Methods}, vol.~73, no.~1, pp.
  69--79, 1997.

\bibitem{gasser1982}
T.~Gasser, P.~B{\"a}cher, and J.~M{\"o}cks, ``Transformations towards the
  normal distribution of broad band spectral parameters of the eeg,''
  \emph{Electroencephalogr. Clin. Neurophysiol.}, vol.~53, no.~1, pp. 119--124,
  1982.

\bibitem{schurmann1996}
J.~Sch{\"u}rmann, \emph{Pattern classification: a unified view of statistical
  and neural approaches}, ser. A Wiley-Interscience publication.\hskip 1em plus
  0.5em minus 0.4em\relax John Wiley \& Sons, 1996.

\bibitem{niedermeyer2005chp9}
E.~Niedermeyer, \emph{The Normal {EEG} of the Waking Adult}.\hskip 1em plus
  0.5em minus 0.4em\relax Lippincott Williams \& Wilkins, 2005, ch.
  Electroencephalography: Basic princ., clin. app., and related fields, p. 167.

\bibitem{Barry}
R.~Barry, A.~Clarke, S.~Johnstone, C.~Magee, and J.~Rushby, ``{EEG} differences
  between eyes-closed and eyes-open resting conditions,'' \emph{Clin.
  Neurophysiol.}, vol. 118, no.~12, pp. 2765--2773, 2007.

\bibitem{bodenmann2009}
S.~Bodenmann, T.~Rusterholz, R.~D{\"u}rr, C.~Stoll, V.~Bachmann, E.~Geissler,
  K.~Jaggi-Schwarz, and H.-P. Landolt, ``The functional val158met polymorphism
  of comt predicts interindividual differences in brain $\alpha$ oscillations
  in young men,'' \emph{J. Neurosci.}, vol.~29, no.~35, pp. 10\,855--10\,862,
  2009.

\bibitem{zietsch2007}
\BIBentryALTinterwordspacing
B.~P. Zietsch, J.~L. Hansen, N.~K. Hansell, G.~M. Geffen, N.~G. Martin, and
  M.~J. Wright, ``Common and specific genetic influences on {EEG} power bands
  delta, theta, alpha, and beta,'' \emph{Biol. Psychol.}, vol.~75, no.~2, pp.
  154 -- 164, 2007. [Online]. Available:
  \url{http://www.sciencedirect.com/science/article/pii/S0301051107000099}
\BIBentrySTDinterwordspacing

\bibitem{saxe2003}
R.~Saxe and N.~Kanwisher, ``People thinking about thinking people: the role of
  the temporo-parietal junction in "theory of mind",'' \emph{Neuroimage},
  vol.~19, no.~4, pp. 1835--1842, 2003.

\bibitem{nolte2004}
G.~Nolte, O.~Bai, L.~Wheaton, Z.~Mari, S.~Vorbach, and M.~Hallett,
  ``Identifying true brain interaction from eeg data using the imaginary part
  of coherency,'' \emph{Clin. Neurophysiol.}, vol. 115, no.~10, pp. 2292--2307,
  2004.

\bibitem{kwiatkowski1992}
D.~Kwiatkowski, P.~C.~B. Phillips, P.~Schmidt, and Y.~Shin, ``Testing the null
  hypothesis of stationarity against the alternative of a unit root,'' \emph{J.
  Econometrics}, vol.~54, no. 1-3, pp. 159--178, 1992.

\bibitem{le_van_quyen2001}
M.~Le~Van~Quyen, J.~Foucher, J.-P. Lachaux, E.~Rodriguez, A.~Lutz,
  J.~Martinerie, and F.~J. Varela, ``Comparison of hilbert transform and
  wavelet methods for the analysis of neuronal synchrony,'' \emph{J. Neurosci.
  Methods}, vol. 111, no.~2, pp. 83--98, 2001.

\bibitem{scarano2012}
G.~Scarano, L.~Forastiere, S.~Colonnese, and S.~Rinauro, ``Reduced polynomial
  classifier using within-class standardizing transform,'' in
  \emph{Communications Control and Signal Processing (ISCCSP), 2012 5th
  International Symposium on}.\hskip 1em plus 0.5em minus 0.4em\relax IEEE,
  2012, pp. 1--4.

\bibitem{chi2010}
Y.~M. Chi, T.-P. Jung, and G.~Cauwenberghs, ``Dry-contact and noncontact
  biopotential electrodes: methodological review,'' \emph{Biomedical
  Engineering, IEEE Reviews in}, vol.~3, pp. 106--119, 2010.

\bibitem{Brigham10}
K.~Brigham and B.~V. Kumar, ``Subject identification from electroencephalogram
  ({EEG}) signals during imagined speech,'' in \emph{Proc. {IEEE} 4th Int.
  Conf. {BTAS}}, 2010, pp. 1--8.

\bibitem{Palaniappan07}
R.~Palaniappan and D.~Mandic, ``Biometrics from brain electrical activity: A
  machine learning approach,'' \emph{IEEE Trans. Pattern Anal. Mach. Intell.},
  vol.~29, no.~4, pp. 738--742, 2007.

\end{thebibliography}
\end{document}